\documentclass[letterpaper,twocolumn,10pt]{sig-alternate-10pt}

\usepackage{wrapfig}
\usepackage{comment}
\usepackage{colortbl}

\usepackage{hyperref}

\usepackage{color,soul}
\usepackage{graphics}
\usepackage{hyperref}
\usepackage{lipsum}
\usepackage{tikz}
\usepackage{enumitem}	
\usepackage{listings} 

\usepackage{booktabs}	
\usepackage{lipsum}

\usepackage{capt-of}	

\usepackage{todonotes}  
\usepackage{subcaption} 

\definecolor{refkey}{rgb}{249,158,26}
\definecolor{labelkey}{rgb}{0,0,1}

\usepackage{cleveref}
\AtBeginDocument{\DeclareCaptionSubType{lstlisting}}
\crefname{sublstlisting}{listing}{listings}
\Crefname{sublstlisting}{Listing}{Listings}

\usepackage{mathrsfs}	
\usepackage{amsfonts}

\usepackage[export]{adjustbox} 

\usepackage{algorithm,algorithmic,amsmath} 
\usepackage{boldline}					

\usepackage{multirow}

\renewcommand{\paragraph}[1]{\vskip 3pt\noindent\textbf{#1 }}	 

  {\begin{list}{$\bullet$}%
     {\setlength{\parsep}{0pt}%
      \setlength{\topsep}{0pt}%
      \setlength{\itemsep}{2pt}}}%
  {\end{list}}
\newcommand\Note[1]{\sethlcolor{yellow} \hl{#1}} 
\newcommand\Noted[1]{} 

\newcommand\xzlNote[1]{\sethlcolor{yellow} \hl{xzl: #1}} 

\definecolor{darkblue}{rgb}{0.0, 0.0, 0.55}
\definecolor{mygreen}{HTML}{ADFF2F}
\definecolor{mylightgray}{gray}{0.8}
\newcommand\jin[2][]{
	\fcolorbox{blue}{white}{\bf\em\color{blue}jin}
	{\small\em\color{blue}{\fontfamily{qhv}\selectfont \underline{#1} #2}}
}

\newenvironment{myitemize}%
  {\begin{itemize}
	[leftmargin=0cm,
		itemindent=.3cm,
		labelwidth=\itemindent,
		labelsep=0pt,
		parsep=0.0pt,
		topsep=0.5pt,
		itemsep=0.5pt,
		align=left]
  }%
  {\end{itemize}}    

\newenvironment{myenumerate}%
  {\begin{enumerate}
	[leftmargin=.cm,itemindent=.5cm,labelwidth=\itemindent,
		labelsep=0pt,
		parsep=1pt,
		topsep=1pt,
		itemsep=3pt,
		align=left]
  }%
  {\end{enumerate}}    

\newcommand\sect[1]{Section~\ref{sec:#1}}	

\newcommand{\code}[1]{\texttt{#1}}

\newcommand{\naive}{\texttt{Naive}}
\newcommand{\cdm}{\texttt{OursM}}
\newcommand{\cdmd}{\texttt{OursMD}}
\newcommand{\cdfull}{\texttt{OursMDS}}

\newcommand{\cloudshim}{DriverShim}
\newcommand{\clientshim}{GPUShim}

\newcommand{\sys}{CODY}
\newcommand{\gpurip}{GPU replay}

\newcommand{\rnr}{record and replay}

\makeatletter
\def\@copyrightspace{\relax}
\makeatother

\begin{document}
\title{Safe and Practical GPU Acceleration in TrustZone}


\author{
	Heejin Park \\
	Purdue ECE \\
	\and
	Felix Xiaozhu Lin \\
	University of Virginia \\

}


\maketitle
\pagestyle{plain}


\begin{abstract}
We present a holistic design for GPU-accelerated computation in TrustZone TEE. 
Without pulling the complex GPU software stack into the TEE, 
we follow a simple approach: 
record the CPU/GPU interactions ahead of time, and replay the interactions in the TEE at run time.
This paper addresses the approach's key missing piece -- the recording environment, which needs both strong security and access to diverse mobile GPUs.
To this end, we present a novel architecture called \sys{}, in which
a mobile device (which possesses the GPU hardware) and a trustworthy cloud service (which runs the GPU software) 
exercise the GPU hardware/software in a collaborative, distributed fashion.  
To overcome numerous network round trips and long delays, 
\sys{} contributes optimizations specific to mobile GPUs: register access deferral, speculation, and metastate-only synchronization. 
With these optimizations, recording a compute workload takes only tens of seconds, which is up to 95\% less than a naive approach; 
replay incurs 25\% lower delays compared to insecure, native execution.

\end{abstract}

\section{Introduction}
\label{sec:intro}




\paragraph{GPU in TrustZone}
Trusted execution environments (TEE) has been a popular facility for secure GPU computation~\cite{graviton,HIX}.
By isolating GPU from the untrusted OS of the same machine, 
it ensures the GPU computation's confidentiality and integrity. 
This paper focuses on GPU computation in TrustZone, the TEE on ARM-based personal devices. 
For these devices, 
in-TEE GPU compute is especially useful, 
as they often run GPU-accelerated ML on sensitive data, e.g. user's health activities, speech audio samples, and video frames. 

\paragraph{GPU stack mismatches TrustZone} 
Towards isolating the GPU \textit{hardware}, TrustZone is already capable~\cite{sanctuary,seCloak}, which is contrast to other TEEs such as SGX. 
The biggest obstacle is the GPU \textit{software} stack, which comprises ML frameworks, a userspace runtime, and a device driver. 
The stack is large, e.g. the runtime for Mali GPUs is an executable binary of 48 MB; it has deep dependency on a POSIX OS, e.g. to run JIT compilation; it is known to contain vulnerabilities~\cite{cudaLeak,CVE-2019-5068,CVE-2020-11179}.
Such a feature-rich stack mismatches the TEE, which expects minimalist software for strong security~\cite{panoply,rubinov16icse,streamBox-TZ}.
Recognizing the mismatch, prior works either transform the GPU stack~\cite{graviton} or the workloads~\cite{schrodinText,slalom,visor}.
They suffer from drawbacks including high engineering efforts and loss of compatibility, as will be analyzed in Section~\ref{sec:bkgnd}. 

\begin{figure}
	\centering
	\includegraphics[width=0.42\textwidth{}]{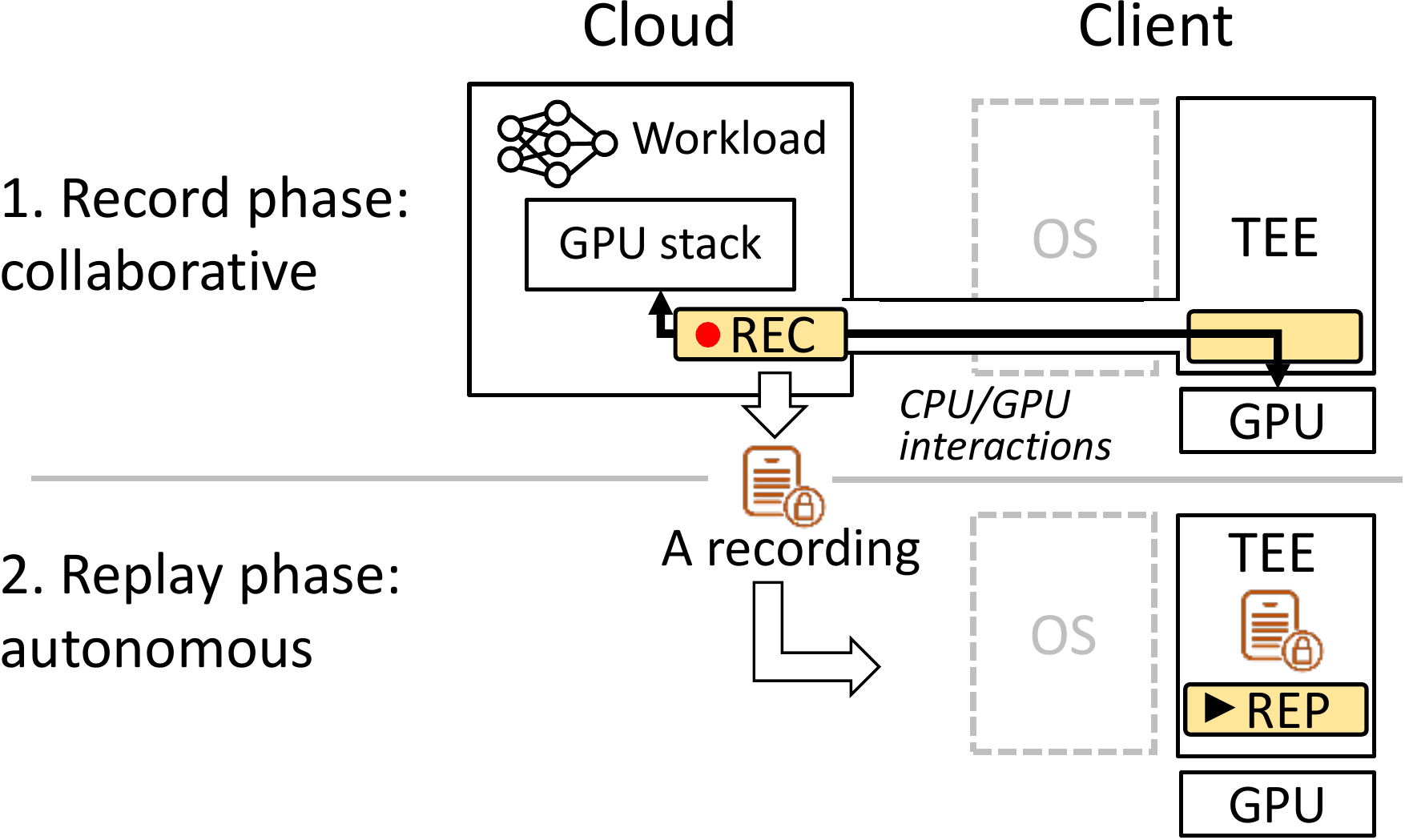}
	\caption{System overview. Shaded components belong to \sys{}}
	\label{fig:overview}
	\vspace{-15pt}		
\end{figure}

\paragraph{Goal \& overall approach}
\textit{Can the TrustZone TEE run GPU-accelerated compute without an overhaul of the GPU stack?} 
To this end, a recent approach called \gpurip{} shows high promise~\cite{tinyStack}.
It executes a GPU-accelerated workload $\mathcal{W}$, e.g. neural network (NN) inference, in two phases. 
(1) In the record phase, developers run $\mathcal{W}$ on a full GPU stack and log CPU/GPU interactions as a series of register accesses and memory dumps. 
(2) In the replay phase, a target program replays the pre-recorded CPU/GPU interactions on new input data without needing a GPU stack. 
\gpurip{} well suites TEE. 
The record phase can be done in a safe environment which faces low threats. 
After record is done \textit{once}, replay can happen within the TEE \textit{repeatedly}. 
The replayer can be as simple as a few KSLoC, has little external dependency, and contains no vulnerabilities seen in a GPU stack~\cite{CVE-2019-14615,CVE-2019-5068,CVE-2020-11179}.
Note that it is crucial to record and replay at CPU/GPU boundary; 
recording at higher levels, e.g. ML framework APIs,
would bloat the TEE with implementation of these APIs.
Yet, a key, unsolved problem is the \textit{recording environment}, 
where the full GPU stack is exercised and CPU/GPU interactions are logged. 
The recording environment must simultaneously (1) enjoy strong security and (2) access the \textit{exact} GPU hardware that will be used for replay. 
These requirements preclude recording on the OS of the same mobile device, as TEE does not trust the OS. 
They also preclude recording on a developer's machine, 
because it can be difficult for developers to predict and possess all GPU hardware models that their workloads \textit{may} execute on. 
Section~\ref{sec:bkgnd} will present details on today's diverse mobile GPUs.

\paragraph{Key idea}
We present a novel approach called collaborative dryrun (\sys{}), 
in which the TEE leverages the cloud for GPU recording. 
As shown in Figure~\ref{fig:overview}, 
a cloud service hosts the GPU software stack without hosting any GPU hardware. 
To record, the TEE on a mobile device (referred to as the ``client'') requests the cloud to run a workload, e.g. NN inference. 
The cloud exercises its GPU stack without executing the actual GPU computation;  
it tunnels all the resultant CPU/GPU interactions between the GPU stack and the physical GPU isolated in the client TEE. 
The cloud logs all the interactions as a \textit{recording} for the workload.
In future execution of the workload on new inputs, the TEE replays the recording on its physical GPU without invoking the cloud service.

\sys{} addresses the needs for a secure, manageable recording environment. 
First, 
unlike mobile devices which face high threats from numerous apps, 
the cloud service runs on rigorously managed infrastructures and only exposes a small attack surface -- authenticated, encrypted communication with the client TEE.  
Importantly, the cloud service never learns the TEE's sensitive data, e.g. ML input and model parameters.
Second, the cloud service accesses the exact, diverse GPU hardware (Figure~\ref{fig:gputrend}) without the hassle of hosting them. 
It is responsible for hosting drivers for the GPU hardware, a task which we will show as practical.
\paragraph{Challenges and Designs}
The main challenge arises from spanning CPU/GPU interactions over the  connection between the cloud and the client. 
A GPU workload generates frequent register accesses (more than 95\% are read), accesses to shared memory, and interrupts. 
If the GPU stack and the GPU hardware were co-located on the same machine, each interaction event takes no more than microseconds; 
since we distribute them over wireless connection, each event will take milliseconds or seconds. 
Forwarding the interactions naively
results in formidable delays, rendering \sys{} unusable. 

To overcome the long delays, we exploit two insights.
(1) The sequence of GPU register accesses consists of many \textit{recurring segments}, corresponding to driver routines repeatedly invoked in GPU workloads, e.g. for job submission and GPU cache flush. 
By learning these segments, the cloud service can predict most register accesses and their outcomes. 
(2) Unlike IO-as-a-service~\cite{io-as-a-service}
which must produce correct results, the cloud only has to extract \textit{replayable interactions} for later actual executions.
%
%
%
With the insights, 
\sys{} automatically instruments the GPU driver code in the cloud to implement the following mechanisms. 

\noindent
(1)
\textit{Register access deferral.} 
While each register access was designed to be executed on the physical GPU synchronously, 
the cloud service queues and commits multiple accesses to the client GPU in a batch, coalescing their network round trips. 
Since register accesses are interleaved with the driver execution in program order, 
the cloud service represents the values of uncommitted register reads as symbols and allows symbolic execution of the driver.
After the register reads are completed by the client GPU, 
the cloud replaces symbolic variables with concrete register values. 

\noindent
(2)
\textit{Register access speculation. }
To further mask the network delay of a commit, 
the cloud service predicts the outcomes of register reads in the commit. 
Without waiting for the commit to finish, the cloud allows the driver to continue execution based on the predicted read values.  
The cloud validates the speculation after the client returns the actual register values. 
In case of misprediction, both the cloud and the client leverage the GPU replay technique to rapidly rollback to their most recent valid states. 

\noindent
(3)
\textit{Metastate-only synchronization. }
Despite physically distributed memories, the driver in the cloud and the client GPU must maintain a synchronized memory view. 
We reduce the synchronization \textit{frequencies} by tapping in GPU hardware events; 
we reduce the synchronization \textit{traffic}
by only synchronizing GPU's metastate -- GPU shaders, command lists, and job descriptions -- while omitting workload data, which constitutes the majority of GPU memory. 
As a result, we preserve correct CPU/GPU interactions 
while forgoing the compute result correctness, a unique opportunity of dryrun. 

\paragraph{Results}
We build \sys{} atop Arm platforms and Mali Bifrost, a popular family of mobile GPUs, and evaluate it on a series of ML workloads. 
Compared to naive approach, 
\sys{} lowers the recording delays by two order of magnitude,
from several hundred seconds to 10 -- 40 seconds;
it reduces the client energy consumption by up to 99\%. 
Its replay incurs 25\% lower delays as compared to insecure, native execution of the workloads. 

\paragraph{Contributions}
We present a holistic solution for GPU acceleration within the TrustZone TEE. 
We address the key missing piece -- a safe, practical recording environment. 
We make the following contributions. 

\begin{myitemize}
\item 
A novel architecture called \sys{}, where the cloud and the client TEE collaboratively exercise the GPU stack for recording CPU/GPU interactions.

\item 
A suite of key I/O optimizations that exploit GPU-specific insights in order to overcome the long network delays between the cloud and the client. 

\item 
A concrete implementation for practicality: 
lightweight instrumentation of the GPU driver; 
crafting the device tree for VMs to probe GPU without hosting the GPU;
a TEE module managing GPU for record and replay. 


\end{myitemize}


\section{Motivations}
\label{sec:bkgnd}

\subsection{Mobile GPUs}
This paper focuses on mobile GPUs which share memory with CPU. 

\paragraph{GPU stack and execution workflow}
As shown in Figure~\ref{fig:workflow}, 
a modern GPU stack consists of ML frameworks (e.g. Tensorflow), a userspace runtime for GPU APIs (e.g. OpenCL), and a GPU driver in the kernel. 

When an app executes ML workloads, it invokes GPU APIs, e.g. OpenCL.
Accordingly, the runtime prepares GPU jobs and input data: 
it emits GPU commands, shaders, and data to the shared memory which is mapped to the app's address space. 
The driver sets up the GPU's pagetables, configures GPU hardware, and submits the GPU job.
The GPU loads the job shader code and data from the shared memory, executes the code, and writes back compute results and job status to the memory.
After the job, the GPU raises an interrupt to the CPU. 
For throughput, the GPU stack often supports multiple outstanding jobs. 



\paragraph{CPU/GPU interactions}
through three channels:

\begin{myitemize}
	\item 
	Registers, for configuring GPU and controlling jobs. 
	\item 
	Shared memory, to which CPU deposits commands, shaders, and data and retrieves compute results.
	Modern GPUs have dedicated pagetables, allowing them to access shared memory using GPU virtual addresses. 
	\item 
	GPU interrupts, which signal GPU job status.
\end{myitemize}

The GPU driver manages these interaction; 
thus it can interpose and log these interactions. 

\begin{figure}
	\centering
	\includegraphics[width=0.45\textwidth{}]{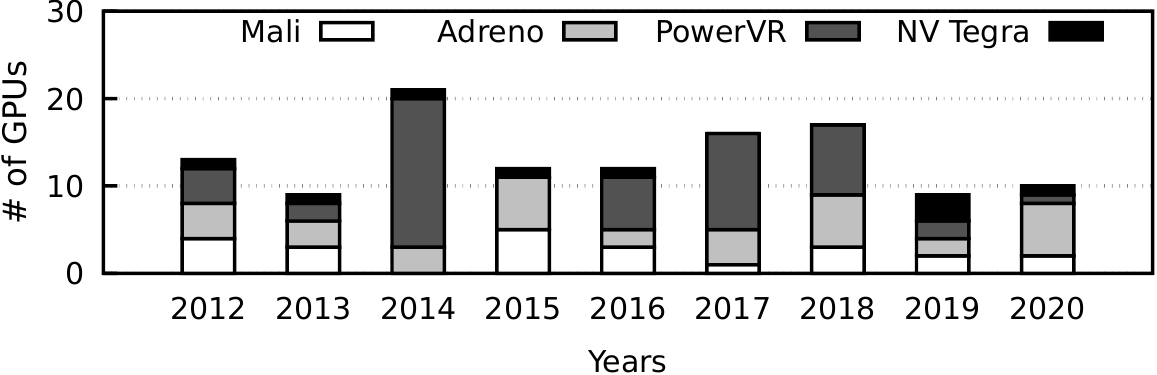}	
	\caption{Numbers of mobile GPU models per year~\cite{gputrend}}
\label{fig:gputrend}
\vspace{-10pt}		
\end{figure}

\subsection{Prior Approaches}
%
%

Our goal is to run GPU compute inside the TrustZone TEE,
for which prior approaches are inadequate. 

\paragraph{Porting GPU stack to TEE}
One approach is to pull the GPU stack to the TEE (``lift and shift'')~\cite{HIX,secDeep}. 
The biggest problem is the clumsy GPU stack: 
the stack spans large codebases (e.g. tens of MB binary code), much of which are proprietary.  
The stack depends on POSIX APIs which are unavailable inside TrustZone TEE. 
For these reasons, 
it will be a daunting task to port proprietary runtime binaries and a POSIX emulation layer, let alone the GPU driver. 
\textit{Partitioning} the GPU stack and porting part of it, as suggested by recent works~\cite{telekine,graviton}, also see significant drawbacks: 
they still require high engineering efforts and sometimes even hardware modification.
The ported GPU code is likely to introduce vulnerabilities to the TEE ~\cite{CVE-2014-1376,CVE-2019-5068,CVE-2019-20577}, bloating the TCB and weakens security. 

\paragraph{Outsourcing}
Another approach is for TEE to invoke an external GPU stack. 
One choice is to invoke the GPU stack in the normal-world OS of the same device. 
Because the OS is untrusted, the TEE must prevents it from learning ML data/parameters and tampering with the result. 
Recent techniques include homomorphic encryption~\cite{slalom,cryptoNet}, ML workload transformation~\cite{occlumency,ternary}, and result validation~\cite{deepAttest}. 
They lack GPU acceleration or support limited GPU operators, often incurring significant efficiency loss.

\begin{figure}
	\centering
	\includegraphics[width=0.45\textwidth{}]{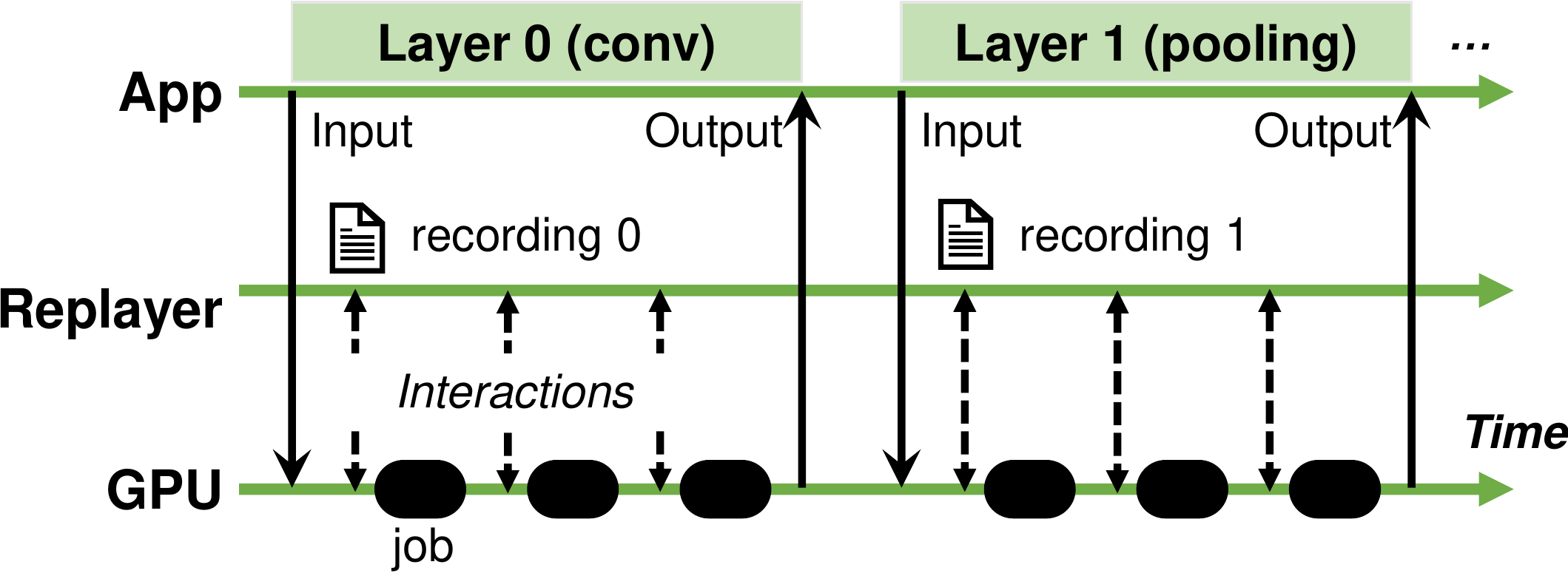}
	\vspace{3pt}		
	\caption{A timeline for replaying NN inference}
	\label{fig:using}
	\vspace{-14pt}		
\end{figure}

\subsection{\gpurip{} in TrustZone}
Unlike prior approaches, 
\gpurip{} provides a new way to execute GPU-accelerated compute~\cite{tinyStack}. 
(1) In the record phase, app developers run their ML workload \textit{once} on a trusted GPU stack;
at the driver level, a recorder logs all the CPU/GPU interactions --
register accesses, GPU memory dumps which enclose GPU commands and shaders, and interrupt events.
These interactions constitute a \textit{recording} for the ML workload. 
(2) In the replay phase, a target app in the TEE supplies new input to the recording. 
The TEE does not need a GPU stack but only a simple replayer (30 KB) for interpreting and executing the logged interactions.

Figure~\ref{fig:using} exemplifies how \gpurip{} works for NN inference. 
To record, developers run the ML inference once and produce a sequence of recordings, one for each NN layer; 
each NN layer invokes multiple GPU jobs, e.g. convolution or pooling. 
To replay, a target ML app executes the recordings in the layer order. 
The granularity of recordings is a developers' choice as the tradeoff between composability and efficiency. 
Alternatively, developers may create one monolithic recording for all the NN layers (not shown in the figure).

\paragraph{Why is \gpurip{} practical?}
(1) An ML workload such as NN often runs pre-defined GPU jobs. 
High-level GPU APIs can be translated to GPU primitives ahead of time; at run time, the workload does not need the stack's dynamic features, e.g. JIT and fine-grained sharing. 
(2) 
An NN often has a static GPU job graph with no conditional branches among jobs. A single record run can exercise all the GPU jobs and record them. 
(3) 
Nondeterministic GPU events can be systematically prevented or tolerated, 
allowing the replayer to faithfully reproduce the recorded jobs. 
For instance, the recorder can serialize GPU job submission and avoid nondeterministic interrupts. 



\subsection{The Problem of Recording Environment}

To apply \gpurip{} to TrustZone, a missing component is the recording environment where the GPU stack is exercised and recordings are produced.
Obviously, the environment should be trustworthy to the TEE.
What is more important, the environment must have access to the GPU hardware that matches the GPU for replay.
Recording with the exact GPU model is crucial. 
In our experience, one shall not even record with a different GPU model from the \textit{same} GPU family, because replay can be broken by subtle hardware differences: 
(1) register values which reflect the GPU's hardware configuration, e.g. shader core count, based on which the JIT compiler generate and optimize GPU shaders; 
(2) encodings of GPU pagetables;
(3) encodings of shared memory, with which GPU communicates its execution status with CPU.


\noindent
\textit{Can recording be done on developers' machines? }
While developers' machines can be trustworthy~\cite{devSecOps}, it would be a heavy burden for the developers to foresee all possible client GPUs and possess the exact GPU models for recording. 
As shown in Figure~\ref{fig:gputrend}, mobile GPUs are highly diverse~\cite{gpuRank}:
today's SoCs see around 80 mobile GPU models in four major families (Apple, PVR, Mali, and Adreno); 
no GPU models are dominating the market; 
new GPU models are rolled out frequently. 


\noindent
\textit{Can recording be done on a ``mobile device farm'' in the cloud?}
While such a device farm relieves developers' burden, 
managing a large, diverse collection of mobile devices in the cloud is tedious if not impractical. 
Not designed to be hosted, mobile devices do not conform to the size, power, heat dissipation requirements of data centers. 
The device farm is not elastic: 
a device can serve one client at a time; planning the capacity and device types is difficult. 
As new mobile devices emerge every few months, the total cost of ownership is high.  

\section{\sys{}}
\label{sec:overview}


\begin{figure}
	\centering
	\includegraphics[width=0.42\textwidth{}]{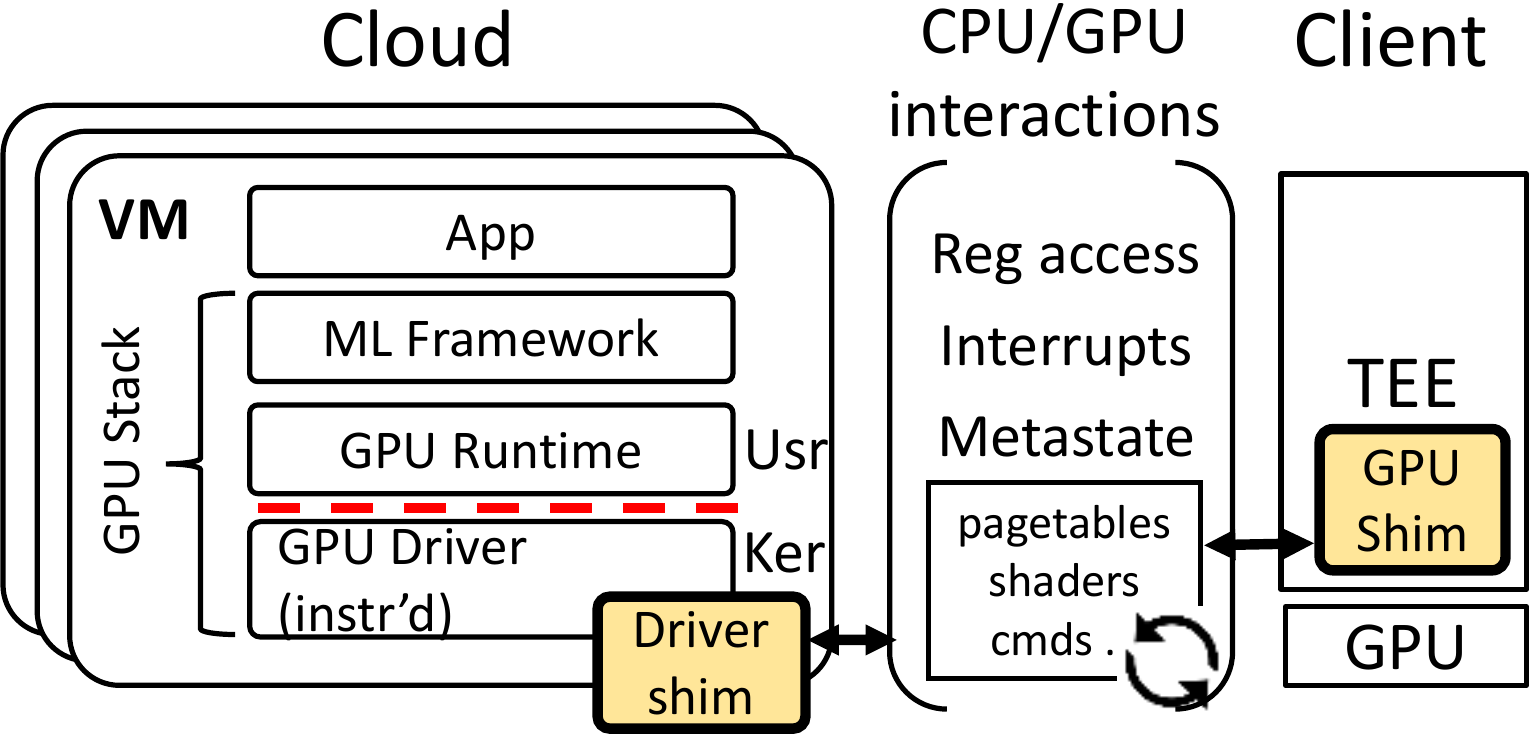}	
	\vspace{3pt}		
	\caption{\sys{}'s online recording. The cloud VM collaboratively runs GPU stack with the client GPU.}
	\label{fig:workflow}
	\vspace{-8pt}		
\end{figure}

We advocate for a new recording environment: 
dryrun the GPU stack in the cloud while using the physical GPUs on the clients.

\subsection{The Approach}



Figure~\ref{fig:workflow} illustrates our approach. 
(1)
Developers write an ML workload as usual, e.g. MNIST inference atop Tensorflow. 
They are oblivious to the TEE, the GPU model, and the cloud service. 
(2)
Before executing the workload for the first time, the client TEE requests the cloud service to dryrun the workload. 
As the cloud runs the GPU stack, it forwards the access to GPU hardware to the client TEE and receives the GPU's response from the latter. 
In the mean time, the cloud records all the CPU/GPU interactions.
(3)
For actual executions of the ML workload, the client TEE replays the recorded CPU/GPU interactions on new input data; 
it no longer involves the cloud. 

Our approach fundamentally differs from remote I/O or I/O-as-a-service~\cite{io-as-a-service}.
Our goal is neither to execute GPU compute in the cloud~\cite{kahawai,clonecloud} (in fact, the cloud has no physical GPUs) nor run the GPU stack \textit{precisely} in the cloud, e.g. for software testing~\cite{charm}.
It is to extract the software's stimuli to GPU and the GPU's response.
This allows \sys{} to skip much communications and optimize the cloud execution. 
\paragraph{Why using the cloud for recording?}
The cloud has the following benefits. 
\begin{myenumerate}
	\item \textit{Rich resources.}
	The cloud can run a GPU stack that is too big to fit in the TEE; 
	it can also host multiple variants of GPU stack, catering to 
	different APIs and frameworks used by ML workloads. 
	
	\item \textit{Secure.}
	The cloud isolates the GPU stack in a safer environment. 
	In contrast to client mobile devices which often run a myriad of apps and face threats such as clickbait and malware, 
	the cloud infrastructure has more rigorous security measures~\cite{trustedCloudComputing,psData}.
	As the dryrun service uses dedicated VMs that only serve authenticated TEEs, the attack surface of the GPU stack is minimized. 
	
	\item \textit{No sensitive data exposed.}
	A client TEE's invocation of dryrun service never gives away its ML model weights or inputs, because recording by design does not need them.
	For this reason, the dryrun service does not have to be hosted in a cloud TEE, e.g. SGX. 
	Section~\ref{sec:eval-security} will present a detailed security analysis. 
\end{myenumerate}

\paragraph{Can the cloud emulate GPUs?}
One may wonder if the cloud operates with software-based GPU emulators~\cite{nomali}, thereby avoid communicating with client GPUs. 
Building such emulators is difficult, 
as it would require precise emulation of GPU interfaces and behaviors.
However, modern GPUs are diverse~\cite{gpuRank};
they often have undisclosed behaviors and interfaces; 
their hardware quirks are not uncommon.


\paragraph{Will the cloud see GPU driver explosion?}
The cloud VMs for dryrun need to install drivers for all GPU models on clients. 
Fortunately, maintaining the drivers will not add much burden, 
as the total number of needed GPU drivers is small. 
A single GPU driver often supports many GPU models of the same family~\cite{midgard-driver,bifrost-driver};
these GPUs share much driver code while differing in register definitions, hardware revisions, and erratum. 
For instance, Mali Bifrost and Qualcomm Adreno 6xx drivers each support 6 and 7 GPUs~\cite{mali-gpuinfo,adreno-gpulist}.
As \sect{env-setup} will show, by crafting the kernel device tree, we can incorporate multiple GPU drivers in one Linux kernel image to be used by the cloud VMs.


\subsection{The \sys{} architecture}
Figure~\ref{fig:workflow} shows the architecture. 
The cloud service manages multiple VM images, each installed with a variant of GPU stack. 
The VM is lean, containing a kernel and the minimal software required by the GPU stack. 
Once launched, a VM is dedicated to serving only one client TEE. 
All the communication between the cloud VM and the TEE is authenticated and encrypted. 

\sys{}'s recorder comprises two shims for the cloud (\cloudshim{}) and for the client TEE (\clientshim{}). 
\cloudshim{} at the bottom of the GPU stack interposes access to the GPU hardware. 
It is implemented by automatic instrumenting of the GPU driver, injecting code to 
register accessors and interrupts handlers. 
\clientshim{}, instantiated as a TEE module, isolates the GPU during recording and prevents normal-world access. 

After a record run, \cloudshim{} processes logged interactions as a recording; it signs and sends the recording back to the client.
To replay, the client TEE loads a recording, verifies its authenticity, and executes the enclosed interactions.
During replay, the TEE isolates the GPU; before and after the replay, it resets the GPU and cleans up all the hardware state. 

\subsection{Challenge: long network delays}
\label{sec:overview:delay}
A GPU stack is designed under the assumption that CPU and GPU co-locate on an on-chip interconnect with sub-microsecond delays. 
\sys{} breaks the assumption by spanning the interconnect over the Internet with tens of ms or even seconds of delays. 
As a result, the GPU driver is blocked frequently.
The GPU driver frequently issues register accesses;
each register access stalls the driver for one round trip time (RTT). 
Taking MNIST inference as an example, the GPU driver roughly issues 2800 register accesses, taking 117 seconds on cellular network.

Long RTTs also make memory synchronization slow. 
\sys{} needs to synchronize the memory views of the driver (cloud) and the GPU (client). 
When they run on the same machine, the driver and the GPU exchange extensive information via shared memory: commands, shader code, and input/output data. 
When the driver and the GPU are distributed, 
maintaining such a shared memory illusion may see prohibitive slowdown. 
As we will show in Section~\ref{sec:mem-sync}, classic distributed shared memory (DSM) misses key opportunity in dryrun.



The long recording delay, often hundreds of seconds shown in \sect{eval}, render \sys{} unusable. 
(1) An ML workload has to wait long before its first execution in TEE. 
(2) During a record run, the TEE must exclusively owns the GPU, 
blocking the normal-world GPU apps for long and hurting the system interactivity. 
(3) The cloud cost is increased, because \sys{} keeps the VMs alive for extended time.
(4) The GPU stack often throws exceptions, because the long delays violate many timing assumptions implicitly made by the stack code.

\section{Hiding Register Access Delays}
\label{sec:reg}


To overcome the long network delays in CPU/GPU interactions, 
we retrofit known I/O optimizations to exploit new opportunities.

\begin{figure}
	\centering
	\includegraphics[width=0.43\textwidth{}]{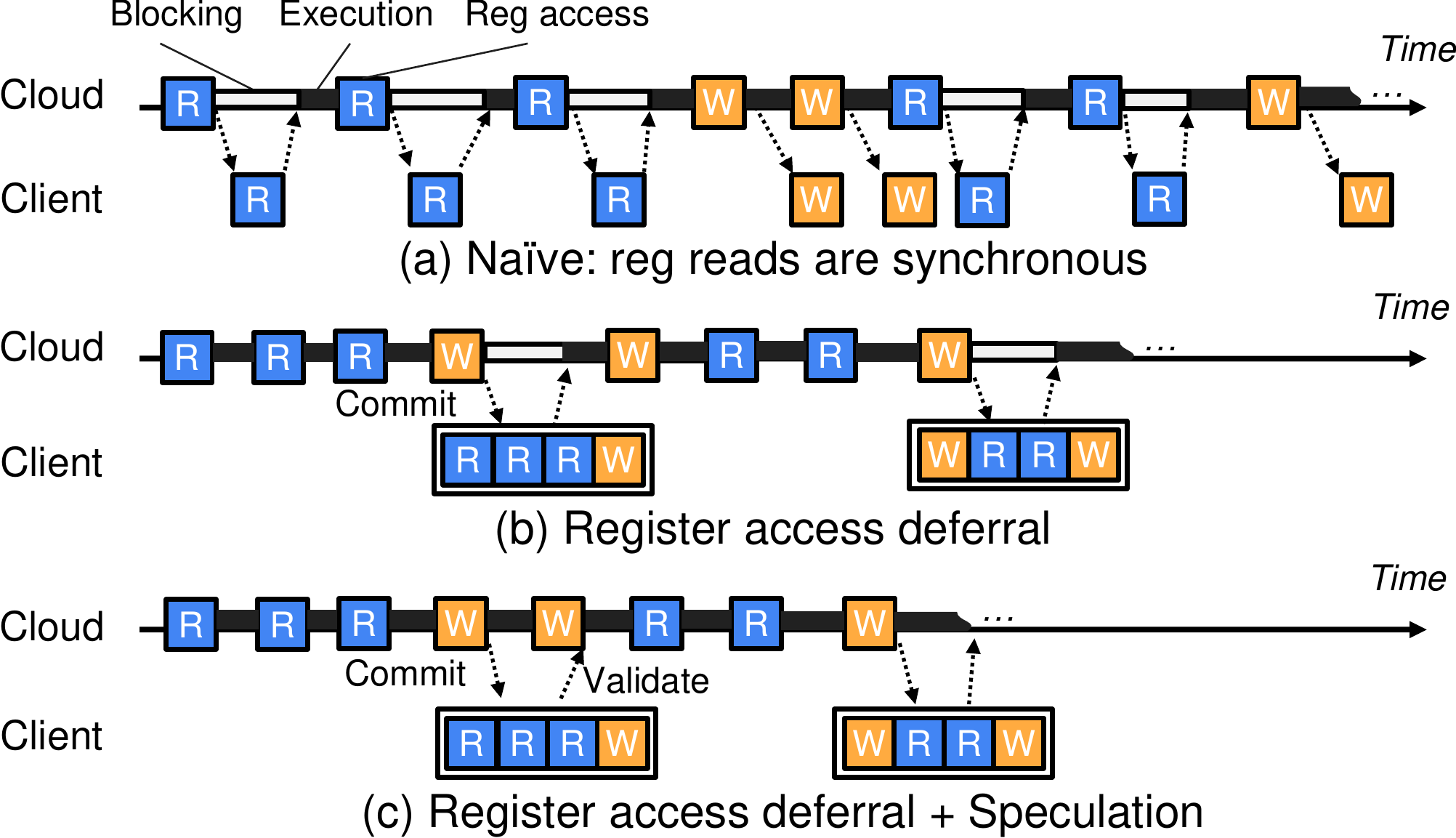}
	\vspace{3pt}		
	\caption{\sys{}'s strategies for hiding long RTTs}
	\label{fig:hiding}
	\vspace{-12pt}		
\end{figure}
\subsection{Register Access Deferral}
\label{sec:reg-defer}

\paragraph{Problem}
By design, a GPU driver weaves GPU register accesses into its instruction stream; 
it executes register accesses and CPU instructions synchronously in program order.
For example in Figure~\ref{fig:hiding}(a), the driver cannot issue the second register access until the first access and the CPU instructions preceding the second register access complete. 
The synchronous register access leads to numerous network round trips.
This is exacerbated by the fact that GPU register accesses are dominated by reads (more than 95\% in our measurement), which cannot be simply buffered as writes. 

\paragraph{Basic idea}
We coalesce the round trips by making register accesses asynchronous:
as shown in Figure~\ref{fig:hiding}(b), 
\cloudshim{} defers register accesses as the driver executes, until the driver cannot continue execution without the value from any deferred register read.
\cloudshim{} then \textit{synchronously} commits all deferred register accesses in a batch to the client GPU. 
After the commit,  
\cloudshim{} stalls the driver execution until the client GPU returns the register access results.


To implement the mechanism, \cloudshim{} injects the deferral hooks into the driver via automatic instrumentation. 
The driver \textit{source code} remains unmodified. 



\paragraph{Key mechanisms for correctness}
First, \cloudshim{} keeps the deferral transparent to the client and its GPU. 
For correctness, the GPU must execute the same sequence of register accesses as if there was no deferral.
The register accesses must be in their \textit{exact} program order, because (1) GPU is stateful and (2) these accesses may have hidden dependencies.
For instance, read from an interrupt register may clear the GPU's interrupt status, which is a prerequisite for a subsequent write to a job register. 
For this reason, \cloudshim{} queues register accesses in their program order.  
It instantiates one queue \textit{per kernel thread}, 
which is important to the memory model to be discussed later. 

\begin{figure}[tb]
	\centering
	\setcaptiontype{lstlisting}

	\begin{subfigure}[b]{0.47\textwidth}   	
		\includegraphics[width=\textwidth]{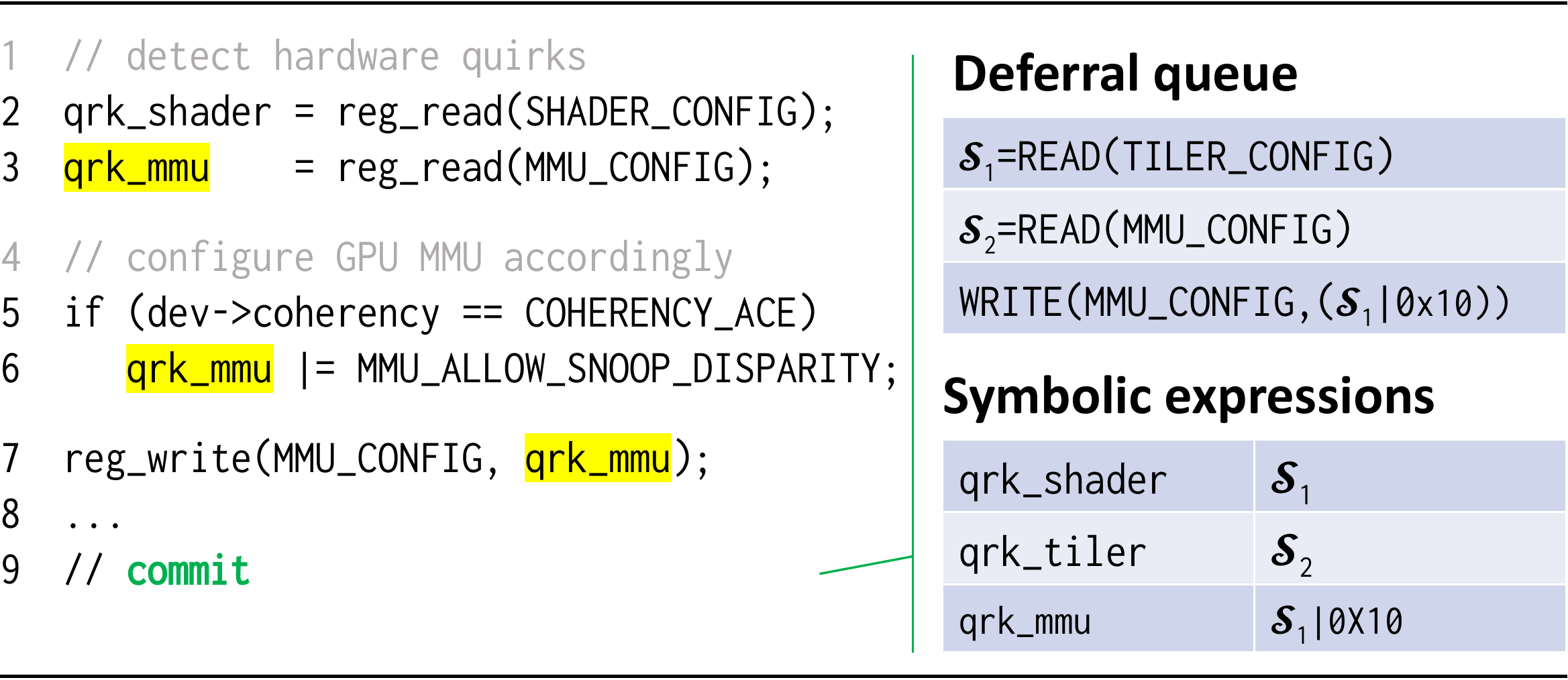}
		\vspace{-15pt}
		\caption{Data dependency}
	\end{subfigure}

	\begin{subfigure}[b]{0.47\textwidth}   	
		\includegraphics[width=\textwidth]{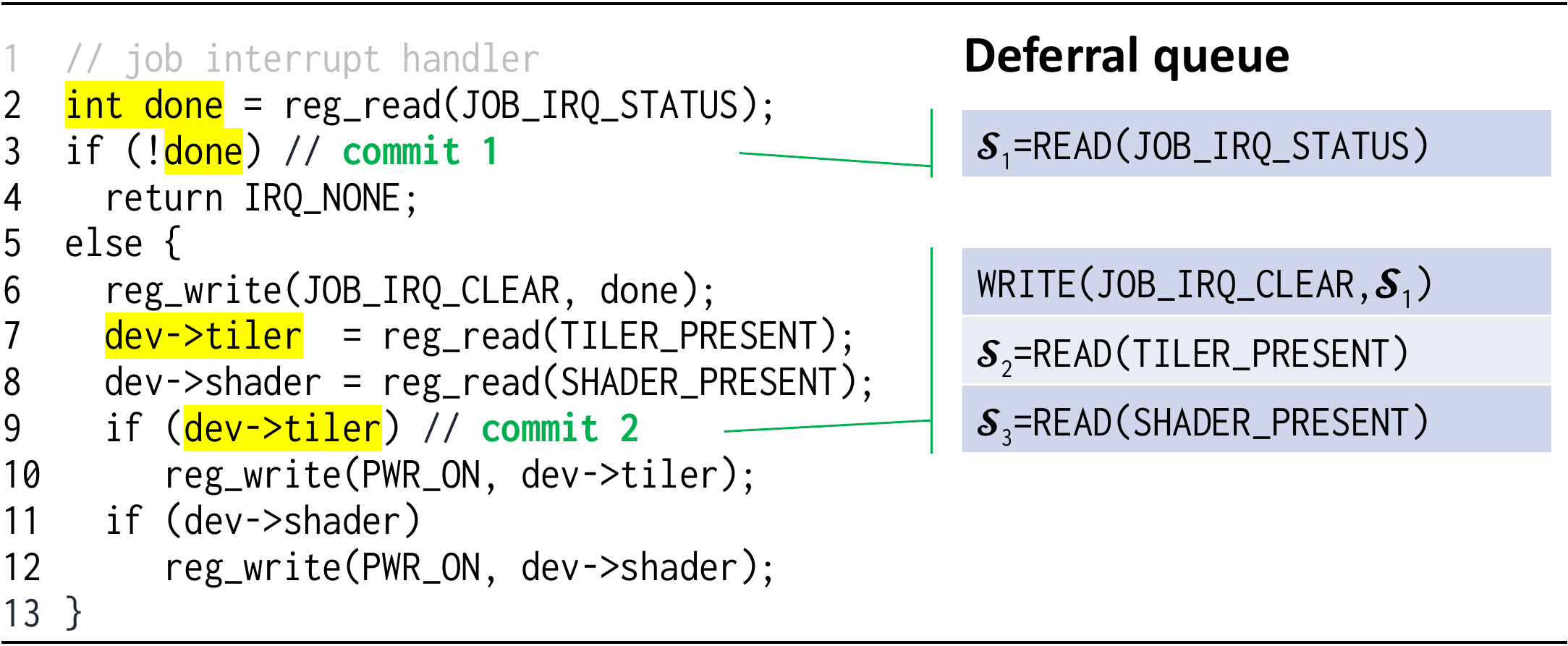}
		\caption{Control dependency (symbolic expressions omitted)}
	\end{subfigure}


	\vspace{3pt}
	\caption{Code examples of data and control dependencies. 
	The register accesses are deferred in the queue; the driver keeps running with symbolic values until commit.}
	\vspace{-10pt}
	\label{list:regio}
\end{figure}

Second, \cloudshim{} tracks data dependencies. 
This is because (1) the driver code may consume values from uncommitted register reads; 
(2) the value of a later register write may depend on the earlier register reads.
Listing~\ref{list:regio} (a) shows examples: 
variable \texttt{qrk\_mmu} depends on the read from register \texttt{MMU\_CONFIG};
the write to \texttt{MMU\_CONFIG} on line 7 depends on the register read on line 3. 
To this end, for each queued register read, \cloudshim{} creates a symbol for the read value and propagates the symbol in subsequent driver execution.
Specifically, a symbol can be encoded in a later register write to be queued, 
e.g. \texttt{reg\_write(MMU\_CONFIG, $\mathcal{S}$|0x10)}, where $\mathcal{S}$ is a symbol. 
After a commit returns concrete register values, 
\cloudshim{} \textit{resolves} the symbols and replaces symbolic expressions in the driver state that encode these symbols. 

Third, \cloudshim{} respects control dependencies. 
The driver control flow may reach a predicate that depends on an uncommitted register read, as shown in Listing~\ref{list:regio} (b), line 3.
\cloudshim{} resolves such control dependency immediately: 
it commits all the queued register accesses including the one pertaining to the predicate.





\lstset{
	language=C++,
	basicstyle=\fontsize{8}{8}\selectfont\ttfamily,
	xleftmargin=5.0ex,
	framexleftmargin=5.0ex,
	frame=tb, 
	breaklines=true,
	captionpos=b,
	numbers=left,
}

%
%
%
%
%
%
%
%
%
\paragraph{When to commit?}
\cloudshim{} commits register accesses when the driver triggers the following events. 

\begin{myitemize}
	\item \textit{Resolution of control dependency}. 
	This happens when the driver execution is about to take a conditional branch that depends on an uncommitted register read. 
	
	\item \textit{Invocations of kernel APIs}, 
	notably scheduling and locking.
	There are three rationales. 
	(1) 
	By doing so, \cloudshim{} safely limits the scope of code instrumentation and dependency tracking to the GPU driver itself; 
	it hence avoids doing so for the whole kernel.
	(2) 
	\cloudshim{} ensures all register reads are completed before kernel APIs that may externalize the register values, e.g. printk() of register values. 
	(3) Committing register accesses prior to any lock operations (lock/unlock) ensures memory consistency, which will be discussed below. 
	
	\item 
	\textit{Driver's explicit delay}, 
	e.g. calling the kernel's delay family of functions~\cite{linux-delays}.
	The drivers often use delays as barriers, assuming register accesses preceding delay() in program order will take effect after delay().
	For example, the driver writes a GPU register to initiate cache flush and then calls delay(), after which the driver expects that the cache flush is completed and coherent GPU data already resides in the shared memory. 
	To respect such design assumptions, \cloudshim{} commits register accesses before explicit delays.
	
\end{myitemize}


\paragraph{Memory consistency for concurrent threads} 
The GPU driver is multi-threaded by design.
Since \cloudshim{} defers register accesses with per-thread queues, 
if a driver thread assigns a symbolic value to a variable $X$, the actual update to $X$ will not happen until the thread commits the corresponding register read. 
What if another thread attempts to read $X$ before the commit? 
Will it read the stale value of $X$? 

\cloudshim{} provides a known memory model of \textit{release consistency}~\cite{comet} to ensure no other concurrent threads can read $X$.  
The memory model is guaranteed by two designs. 
(1) 
Given that the Linux kernel and drivers have been thoroughly scrutinized for data race~\cite{kscan}, 
a thread always updates shared variables (e.g. $X$) with necessary locks, which prevent concurrent accesses to the variables.
(2)
\cloudshim{} always commits register accesses before the driver invokes unlock APIs, i.e. 
a thread commits register accesses before releasing any locks. 
As such, the thread must have updated the shared variables with concrete values before any other threads are allowed to access the variables. 

\paragraph{Optimizations}
To further lower overhead, 
we narrow down the scope of register access deferral. 
We exploit an observation: GPU register accesses show high locality in the driver code: 
tens of ``hot'' driver functions issue more than 90\% register accesses.
These hot functions are analogous to compute kernels in HPC applications. 

To do so, we obtain the list of hot functions via profiling offline.
We run the GPU stack, trace register accesses, and bin them by driver functions. 
At record time, \cloudshim{} only defers register accesses within these functions. 
When the driver's control flow leaves one hot function but not entering another, \cloudshim{} commits queued register accesses. 
Note that 
(1) the choices of hot functions are for optimization and do not affect driver correctness, 
as register accesses outside of hot functions are executed synchronously; 
(2) 
profiling is done \textit{once} per GPU driver, hence incurring low effort.

\subsection{Speculation}
\label{sec:speculation}

\paragraph{Basic idea}
Even with deferred register accesses, 
each commit is still synchronous taking one RTT (Figure~\ref{fig:hiding}(b)). 
\cloudshim{} further makes \textit{some} commits asynchronous to hide their RTTs. 
The idea is shown in Figure~\ref{fig:hiding}(c): 
rather than waiting for a commit $\mathcal{C}$ to complete,  
\cloudshim{} predicts the values of all register reads enclosed in $\mathcal{C}$ and continues driver execution with the predicated values; 
later, when $\mathcal{C}$ completes with the actual read values, 
\cloudshim{} validates the predicated values: 
it continues the driver execution if the all predictions were correct; 
otherwise, 
it initiates a recovery process, 
as will be discussed below. 
Misprediction incurs performance penalty but does not violate correctness. 



\paragraph{Why are register values predictable?}
The efficacy of speculation hinges on predictability of register values. 
Our observation is that 
the driver issues \textit{recurring segments} of register accesses, to which the GPU responds with identical values most of time. 
Such  segments recur within a workload (e.g. MNIST inference) and across workloads (e.g. MNIST and AlexNet inferences). 

Why recurring segments? We identify the following common causes. 
(1) Routine GPU maintenance.
For instance, before and after each GPU job, the driver flushes GPU's TLB/cache. 
The sequences of register accesses and register values (e.g. the final status of flush operations) repeat themselves. 
(2) Repeated GPU state transitions. 
For instance, each time an idle GPU wakes up, 
the driver exercises the GPU's power state machine, 
for which the driver issues a fixed sequence of register writes (to initiate state changes) and reads (to confirm state changes).
(3) Repeated hardware discovery. 
For instance, 
during its initialization, the driver probes GPU hardware capabilities by reading tens of registers. 
The register values remain the same as the hardware does not change. 

\paragraph{When to speculate?}
Not all register accesses belong to recurring segments. 
To minimize misprediction, \cloudshim{} acts \textit{conservatively}, only making prediction when the history of commits shows high confidence. 

When \cloudshim{} is about to make a commit $\mathcal{C}$, it looks up the commit history at the same driver source location.
It considers the most recent $k$ historical commits that enclose the same register access sequence as $\mathcal{C}$:
if all the $k$ historical commits have returned identical sequences of register read values,  \cloudshim{} uses the values for prediction; 
otherwise, \cloudshim{} avoids speculation for $\mathcal{C}$, executing it synchronously instead.
$k$ is a configurable parameter controlling the \cloudshim{}'s confidence that permits prediction. We set $k=3$ in our experiment. 

\paragraph{How does driver execute with predicted values?}
Based on predicted register values, 
the GPU driver may mutate its state and take code branches;
\cloudshim{} may make a new commit without waiting for outstanding commits to complete. 
To ensure correctness, \cloudshim{} stalls the driver execution until all outstanding commits are completed and the predictions are validated, when the driver is about to externalize \textit{any} kernel state, e.g. calling printk() on a variable.
This condition is simple, not differentiating if the externalized state depends on predicted register values. 
As a result, checking the condition is trivial: \cloudshim{} just intercepts a dozen of kernel APIs that may externalize kernel state.
\cloudshim{} eschews from fine-grained tracking of data and control dependencies throughout the whole kernel. 

\textit{Optimization:}
Only checking the above condition has a drawback: in the event of misprediction, both the driver and the GPU have to roll back to valid states, 
because both may have executed based on mispredicted register values. 
Listing~\ref{list:regio} (b) shows an example: 
if the read of \code{JOB\_IRQ\_STATUS} (line 9) is found to be mispredicted after the second commit (line 10), 
the driver already contains incorrect state (in \code{dev}) and the GPU has executed incorrect register accesses (e.g. write to \code{JOB\_IRQ\_CLEAR}). 

To this end, 
\cloudshim{} can relieve the client GPU from rollback in case of misprediction. 
It does so by prevent spilling speculative state to the client. 
%
Specifically,
\cloudshim{} \textit{additionally} stalls the driver before committing register accesses that themselves are speculative, i.e. having dependencies on predicted values. 
For example, in Listing~\ref{list:regio} (b), the second commit must be stalled if the first is yet to complete, because the second commit consists of register accesses (\code{JOB\_IRQ\_CLEAR} and \code{TILER/SHADER\_PRESENT}) that casually depend on the outcome of the first commit. 
To track speculative register accesses, \cloudshim{} \textit{taints} the predicted register values and follow their data/control dependencies in the driver execution. 
In the above example, when the driver takes a conditional branch based on a speculative value (line 3), \cloudshim{} taints all updated variable and statements on that branch to be speculative, e.g. \code{dev->tiler}.
For completeness, the taint tracking applies to any kernel code invoked by the driver. 

\paragraph{How to recover from misprediction?}
When \cloudshim{} finds an actual register value different from what was predicated,
the GPU stack and/or the GPU should restore to valid states. 
We exploit the GPU replay technique~\cite{tinyStack} for both parties to restart and fast-forward \textit{independently}. 
To initiate recovery, \cloudshim{} sends the client the location of the mispredicted register access in the interaction log. 
Then both parties restart and replay the log up to the location. 
In this process, 
\clientshim{} feeds the recorded stimuli (e.g. register writes) to the physical GPU; 
\cloudshim{} feeds the recorded GPU response (e.g. register reads and interrupts) to the GPU stack. 
Because both parities need no network communication, the recovery takes only a few seconds, as will be evaluated in Section~\ref{sec:eval-choice}.

\subsection{Offloading polling loops} 
\label{sec:off-polling}

\begin{figure}[t]
	\centering
	\setcaptiontype{lstlisting}
%
	
	\includegraphics[width=.42\textwidth]{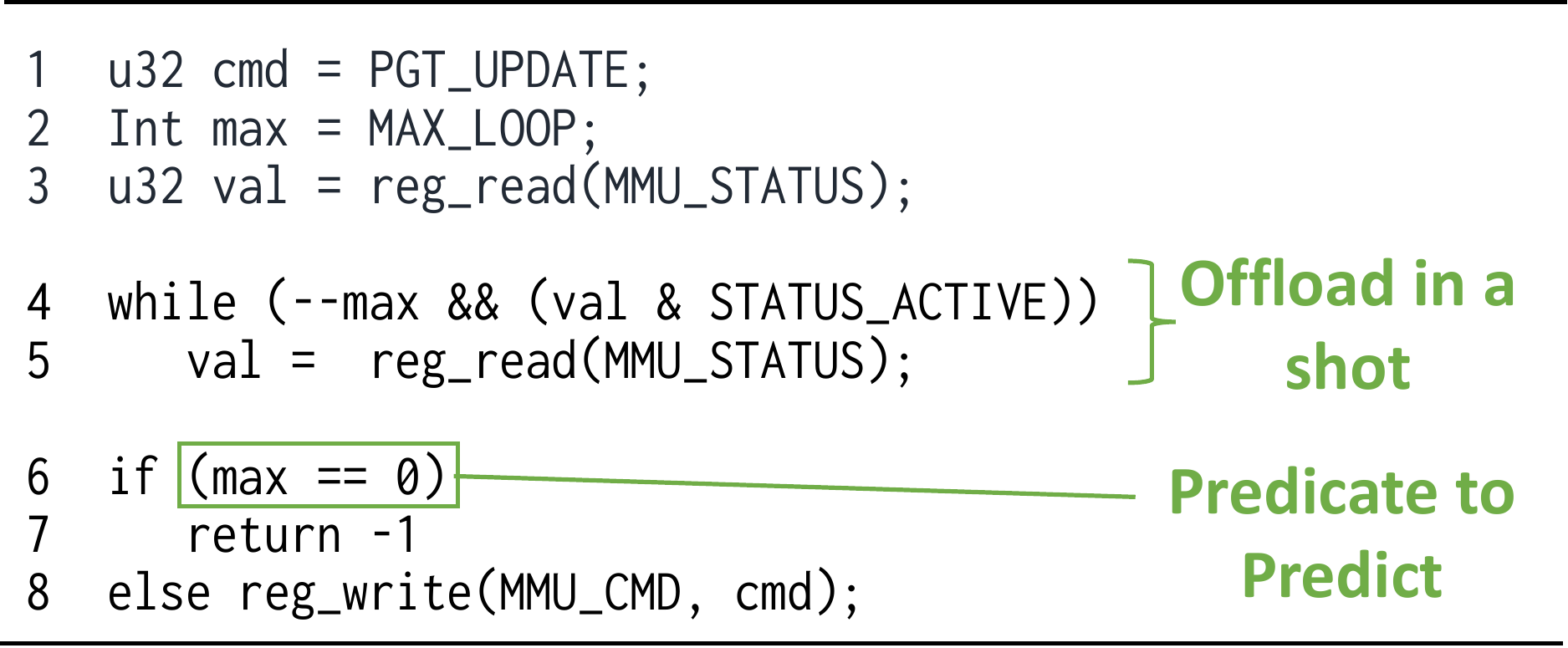}	
	\vspace{3pt}
	\caption{Code example of a polling loop}
	\vspace{-13pt}
	\label{list:polling}
\end{figure}

A GPU driver often invokes polling loops, e.g. to busy wait for register value changes as shown in Listing~\ref{list:polling}.
Polling loops contribute a large fraction of register accesses; 
they are a major source of control dependencies.


\lstset{
	language=C++,
	basicstyle=\fontsize{8}{8}\selectfont\ttfamily,
	xleftmargin=5.0ex,
	framexleftmargin=5.0ex,
	frame=tb, 
	breaklines=true,
	captionpos=b,
	numbers=left,
	tabsize=2,
}

\paragraph{Problem}
Naive execution of a polling loop incurs multiple round trips, rendering
the aforementioned techniques ineffective. 
(1) Deferring register access does not benefit much, because each loop iteration generates control dependency and requests a synchronous commit. 
(2) Speculation on a polling loop is difficult: 
by design above,
\cloudshim{} must predict the iteration count before the terminating condition is met, 
which often depends on GPU timing (e.g. a GPU job's delay) and is nondeterministic in general. 

\paragraph{Observations}
Fortunately, most of polling loops are simple, meeting the following conditions.

\begin{myitemize}
\item Register accesses in the loop are \textit{idempotent}: 
the GPU state is not be affected by re-execution of the loop body.

\item The iteration count has only local impact:
the count is a local variable and does not escape the function enclosing the loop. 
The count is evaluated with some simple predicates, 
e.g.
\texttt{(count<MAX)}.

\item The addresses of kernel variables referenced in a loop are determined prior to the loop, i.e. the loop itself does not compute these addresses dynamically. 

\item 
The loop body does not invoke kernel APIs that have external impact, e.g. locking and printk(). 

\end{myitemize}

Simple polling loops allow optimizations as will be discussed below.
\cloudshim{} uses static analysis to find all of them in the GPU driver.
Complex polling loops that misfit the definition above are rare;
\cloudshim{} just executes them without optimizations.


\paragraph{Solution}
\cloudshim{} executes simple polling loops as follows.
(1) \textit{Offloading.}
\cloudshim{} commits a loop in a shot to the client GPU, 
incurring only one RTT. 
To do so, 
\cloudshim{} offloads a copy of the loop code as well as all variables to be referenced in the loop. 
\clientshim{} runs the loop and returns updated variables. 
Offloading respects release memory consistency as described in Section~\ref{sec:reg-defer}, 
because accesses to shared variables inside the loop must be protected with locks and the loop itself does not unlock.
(2) \textit{Speculation.}
\cloudshim{} further masks the RTT in offloading a loop. 
Rather than predicting the exact iteration count (e.g. the final value of \code{max} in Listing~\ref{list:polling}), \cloudshim{} extracts and predicts the \textit{predicate} on the iteration count, e.g. \code{(max?=0)}, which is more predictable.
When the client returns the actual iteration count, \cloudshim{} evaluates the predicate in order to validate the prediction. 




\section{Memory Synchronization}
\label{sec:mem-sync}

\begin{figure}
	\centering
	\includegraphics[width=0.45\textwidth{}]{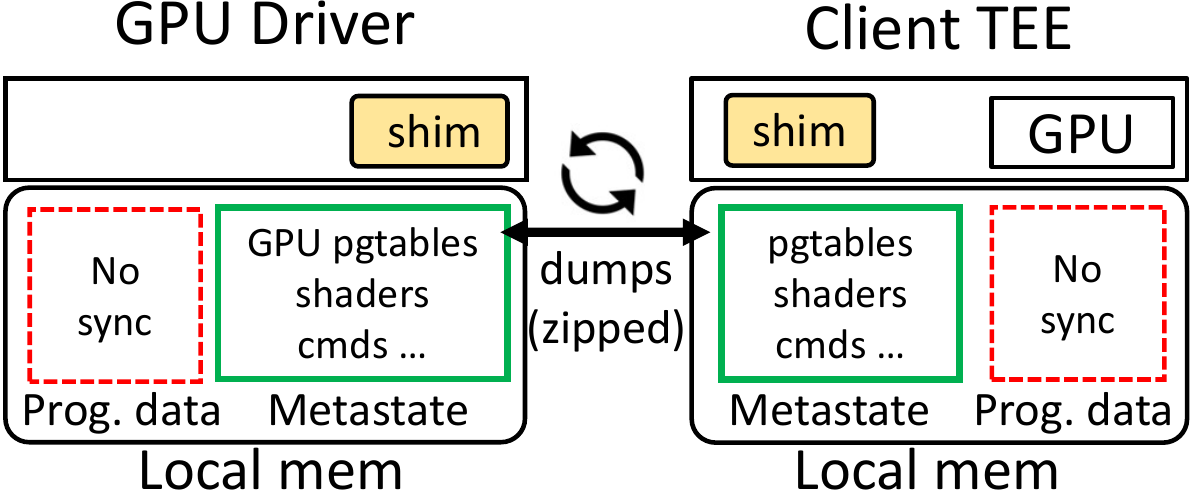}	
	\vspace{3pt}		
	\caption{Selective memory synchronization of GPU metastate but not program data}
\label{fig:mmsync}
\vspace{-12pt}		
\end{figure}

\paragraph{Problem}
While the driver (cloud) and the GPU (client) run on their own local memoriess, 
we need to synchronize the view of shared memory between them as in Figure~\ref{fig:mmsync}.
Memory synchronization has been a central issue in distributed execution\cite{clonecloud,comet,rio, charm}. 
A proven approach is relaxed memory consistency: one node pushes its local memory updates to other nodes only when the latter nodes are about to see the updates. 
Accordingly, prior systems choose synchronization points based on 
program behaviors, 
e.g. synchronizing thread-local memory at the function call boundary~\cite{clonecloud} or synchronizing shared memory of a data-race free program at the lock/unlock operations~\cite{comet}. 

Unlike these prior systems, the memory sharing protocol between CPU and GPU is never explicitly defined.
For example, they never use locks. 
From our observations, we make an educated guess that CPU and GPU write to disjoint memory regions and order their memory accesses by \textit{some} register accesses and \textit{some} driver-injected delays.
However, it would be error-prone to build \sys{} based on such brittle, vague assumptions. 

\paragraph{Approach}
Our idea is to constrain the GPU driver behaviors so that we can make \textit{conservative} assumptions for memory synchronization. 
To do so, we configure the driver's job queue length to be 1, which effectively serializes the driver's job preparation and the GPU's job execution. 
Such a constraint has been applied in prior work and shows minor overhead, because individual GPU compute jobs are sizable~\cite{tinyStack}. 
With the constraint, the driver prepares GPU jobs (and accesses the shared memory) only when the GPU is idle; the GPU is executing jobs (and accesses the memory) only when the driver is idle. 
As a result, we maintain an invariant: 

\textit{The driver and the client GPU will never access the shared memory simultaneously}. 



\paragraph{When to synchronize?}
The cloud and client synchronize when the GPU is about to become busy or idle:


\begin{myitemize}
\item \textit{Cloud $\Rightarrow$ client.} \vspace{1pt}
Right before the register write that starts a new GPU job, \cloudshim{} dumps kernel memory that the driver allocates for the GPU and sends it to the client.
The memory dump is consistent: at this moment, the GPU driver has emitted and flushed all the memory state needed for the new job, and has updated the GPU pagetables for mapping the memory state.

\item \textit{Client $\Rightarrow$ cloud.} \vspace{1pt}
Right after the client GPU raises an interrupt signaling job completion, \clientshim{} forwards the interrupt and uploads its memory dump to the cloud.
The memory dump is also consistent: at this moment the GPU must have written back the job status and flushed job data from cache to local memory. 
Specifically, the GPU cache flush action is either prescribed in the command stream~\cite{bifrost-cache-flush} or requested at the beginning of the interrupt handler~\cite{vc4-cache-flush}. 
\end{myitemize}

To further safeguard the aforementioned invariant, we implement  continuous validation. 
After \cloudshim{} sends its memory dump to the client, it unmaps the dumped memory regions from CPU and disables DMA to/from the memory. 
As such, any spurious access to the memory region will be trapped to \cloudshim{} as a page fault and reported as an error. 
In the same fashion, \clientshim{} unmaps the shared memory from the \textit{GPU}'s pagetable when the GPU becomes idle; any spurious access from GPU will be trapped and reported. 



\paragraph{What to synchronize?}
As shown in Figure~\ref{fig:mmsync},
we minimize the amount of memory transfer with the following insight:
\textit{for recording, it is sufficient to synchronize only the GPU metastate in memory}, including GPU commands, shader code, and job descriptors. 
Synchronizing program data, including input/output and intermediate GPU buffers, is unnecessary.
This is effective as program data constitutes most of GPU memory footprint. 


How to locate metastate in the shared memory, given that the memory layout is often proprietary? 
We implement a combination of techniques.
(1) Some GPU page tables have permission bits which suggest the usage of memory pages. 
For instance, the Mali GPUs map metastate as \textit{executable} because the state contains GPU shader code~\cite{bifrost-mem-hint}. 
(2) For GPU hardware lacking permission bits, \sys{} infers the usage of memory regions from IOCTL() flags used by ML workloads to map these regions.
For instance, a region mapped as readonly cannot hold GPU commands, 
because the GPU runtime needs the write permission to emit GPU commands. 
%
%
(3) If the above knowledge is unavailable, the \cloudshim{} simply replaces an ML workload's inputs and parameters as zeros. 
Doing so will result in abundant zeros in the GPU's program data, making memory dumps highly compressible.

Atop selective memory synchronization, we apply standard compression techniques.
Both shims use range encoding to compress memory dumps; 
each shim calculates and transfers the deltas of memory dumps between consecutive synchronization points.  

\section{Implementations}
\label{sec:env-setup}

\paragraph{Platforms}
We implement the \sys{} prototype on the following platforms.
The cloud service runs on Odroid C4, an Arm board with 4 Cortex-A55 cores. 
The client runs on Hikey960 which has 4 Cortex-A73 and A53 cores, and a Mali G71 MP8 GPU.
Our choice of Arm processors for the cloud is for engineering ease rather than a hard requirement; 
the cloud service can run on x86 machines with binary translation~\cite{charm}. 

The cloud service runs Debian 9.4 (Linux v4.14) with a GPU stack composed of a ML framework (ACL v20.05 \cite{acl}), a runtime (\code{libmali.so}), and a driver (Mali Bifrost r24~\cite{bifrost-driver}).
Under the cloud service, KVM-QEMU (v4.2.1) runs as the VM hypervisor.
The client runs Debian 9.13 (Linux v4.19) and OPTEE (v3.12) as its TEE.

\paragraph{\cloudshim{}}
We build our code instrumentation tool as a Clang plugin.
For static analysis and code manipulation,
the plugin traverses the driver's abstract syntax tree (AST). 
With the Clang/LLVM toolchain~\cite{clang}, our tool compiles the GPU driver and links it against \cloudshim{}. 
By limiting the scope to the hot driver functions in the Mali GPU driver~(\S\ref{sec:reg-defer}), our instrumentation tool processes 19 functions in total. 
The instrumentation itself incurs negligible overhead. 
We implement \cloudshim{} as a kernel module ($\sim$1K SLoC) to be invoked by the instrumented driver code;
the module performs dependency tracking, commit management, and speculation, as described in \sect{reg} and \ref{sec:mem-sync}.


\cloudshim{} communicates with the client via TCP-based messages in our custom formats. 
To avoid potential timeout due to network communications,
we add a fixed delay (e.g. 3 seconds) to all the timeout values in the driver.
We prepare and install GPU devicetrees in the cloud VM, so the GPU stack can run transparently even a physical GPU is not present~\cite{charm}. 
To support multiple GPU types, 
we implement a mechanism for the cloud service to load per-GPU devicetree when a VM boots. 
As a result, a single VM image can incorporate multiple GPU drivers, which are dynamically loaded depending on the specific client GPU model.

\paragraph{\clientshim{}}
We build \clientshim{} as a TEE module. 
Following the TrustZone convention, 
\clientshim{} communicates with the cloud using the GlobalPlatform APIs implemented by OPTEE~\cite{globalplatform-tee-guide}. 
The communication is authenticated and encrypted by SSL 3.0 with the TEE, 
before it forwarded through the normal-world OS. 

By design, the client's trusted firmware dynamically switches the GPU between the normal world and the TEE with a configurable TrustZone address space controller (TZASC)~\cite{seCloak}.
Yet, our client platform (Hikey960) has a proprietary TZASC which lacks public documentation~\cite{vtz}. 
We workaround this issue by statically reserving memory regions for GPU
and mapping the memory regions and GPU registers to the TEE.  

We modify the secure monitor to route the GPU's interrupts to the TEE.
\clientshim{} forwards the interrupts to \cloudshim{} for handling.
We avoid interrupt injection to the VM hypervisor and keep it unmodified. 




To bootstrap the GPU, the client TEE may need to access SoC resources not managed by the GPU driver, e.g. power/clock for GPU. 
While the TEE may invoke related kernel functions in the normal-world OS via RPC~\cite{charm},
we protect these resources inside the TEE as did in prior work for stronger security~\cite{seCloak}.

\section{Evaluation}
\label{sec:eval}


\begin{figure*}
	\centering
    \begin{subfigure}[b]{0.483\textwidth}   	
		\includegraphics[width=\textwidth]{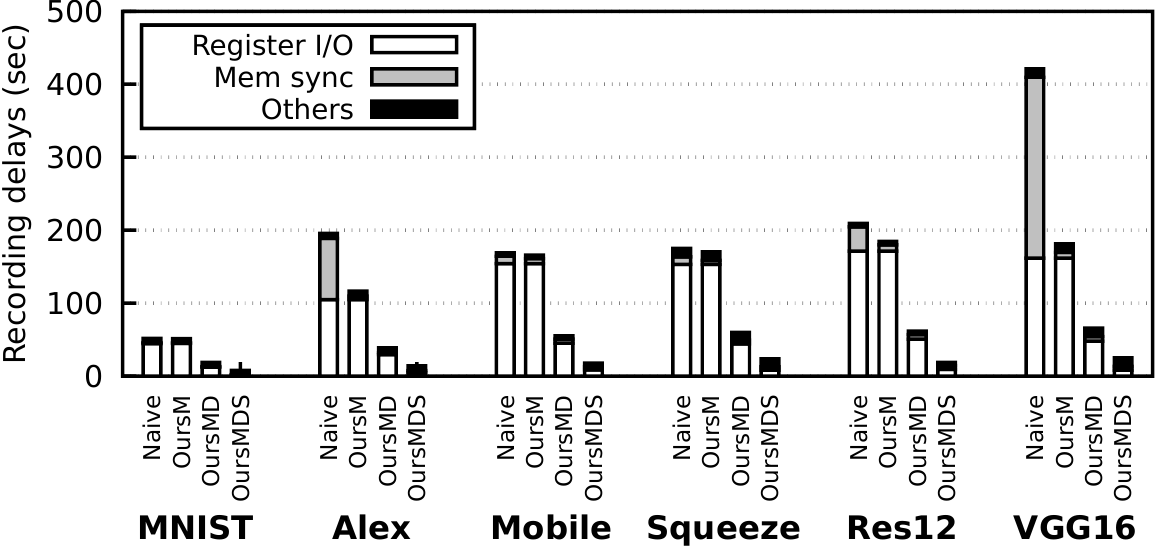}
		\vspace{-15pt}		
		\caption{Recording with WiFi conditions (RTT: 20ms, BW: 80Mbps)}
	\end{subfigure}
	\hfill
	\begin{subfigure}[b]{0.49\textwidth}
		\includegraphics[width=\textwidth]{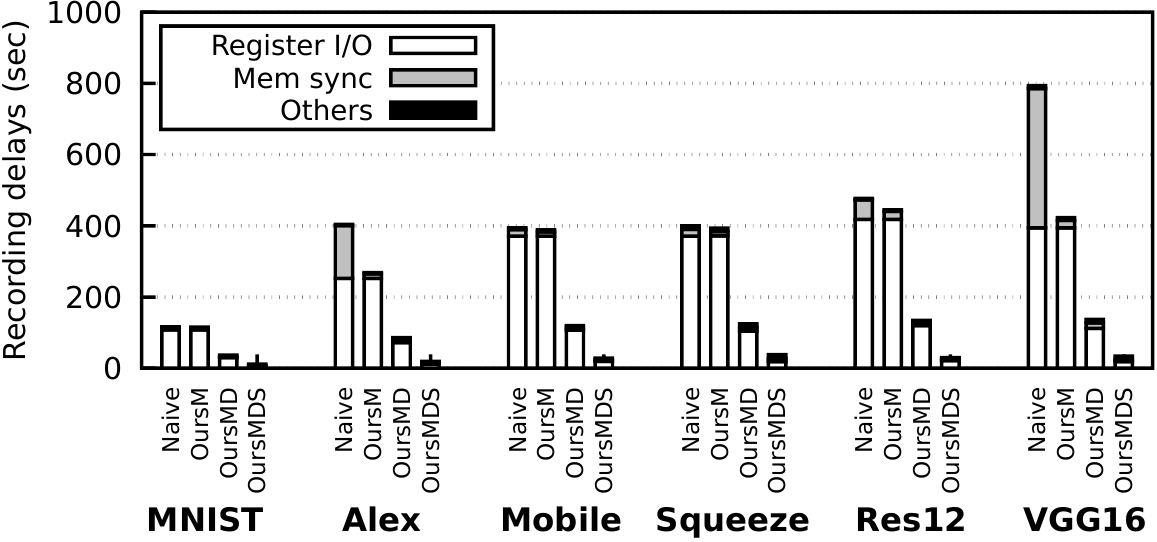}
		\vspace{-15pt}		
		\caption{Recording with cellular conditions (RTT: 50ms, BW: 40Mbps)}
	\end{subfigure}
	

	\caption{Recording delays of \sys{}(\cdfull{}) are significantly lower than other versions}
	\label{fig:recording-delay}
\end{figure*}
The evaluation answers the following questions.

\begin{myitemize}
	\item
	Is \sys{} secure against attacks?~(\S~\ref{sec:eval-security})
	
	\item
	What are the delays of \sys{}?~(\S~\ref{sec:eval-perf})
	
	\item
	Are \sys{}'s optimizations significant?  (\S~\ref{sec:eval-choice})
	
	\item 
	What is the energy implication of \sys{}? (\S\ref{sec:eval:energy})
\end{myitemize}

	

%
%

\subsection{Security Analysis}
\label{sec:eval-security}
\paragraph{Threat model}
We trust the cloud service, assuming its GPU stack is being attested~\cite{trustedCloudComputing,psData}. 
We trust the client's TEE and hardware but not its OS. 
We consider two types of adversaries: 
(1) a local, privileged adversary who controls the client OS; 
(2) a network-level adversary who can eavesdrop the cloud/client communications during recording.

\paragraph{Integrity}
\sys{}'s \textit{recording integrity} is collaboratively ensured by (1) the trusted cloud service, (2) the client's TrustZone hardware, and (3) the encrypted cloud/client communication. 
In particular, \clientshim{} locks the GPU MMIO region during recording, preventing any local adversary from tampering with GPU registers or shared memory. 
\sys{}'s \textit{replay integrity} is ensured by the TrustZone hardware. 
Since the replayer only accepts recordings signed by the cloud, it exposes no additional attack surface to adversaries. 

\paragraph{Confidentiality}
\sys{}'s \textit{recording} never leaks program data from TEE, e.g. ML model parameters or inputs, since recording does not require such data. 
It however may leak some information about the ML workload, 
as the workload \textit{code} such as GPU shaders moves through the network. 
Although the network traffic is encrypted, it may nevertheless leak workload information, e.g. NN types, via side channels. 
Such side channels can be mitigated by orthogonal solutions~\cite{telekine,opaque}.

Since \textit{replay} is within the client TEE and requires no client/cloud communication, its data confidentiality is given by TrustZone. 
We notice TrustZone may leak data to local adversaries via hardware side channels, which can be mitigated by existing solutions~\cite{TruSpy,ARMageddon}.

\paragraph{Availability}
Like any cloud-based service, 
\textit{recording} availability of \sys{} depends on network conditions and the cloud availability, which are vulnerable to DDoS attacks. 
Its \textit{replay} availability is at the same level of the TrustZone TEE, given the GPU power is managed by the TEE not the OS~\cite{seCloak}.

\subsection{Performance}
\label{sec:eval-perf}

\paragraph{Methodology}
As shown in Table~\ref{tab:model_desc}, we test \sys{} on inference with 6 popular NNs running atop ARM Compute Library~\cite{acl}.
We measure \sys{}'s recording delay under two network conditions as controlled by NetEm~\cite{netEm}:
i) WiFi-like (20 ms RTT, 80 Mbps) and ii) cellular-like (50 ms RTT, 40 Mbps)~\cite{ExLL}.
The hardware platform is described in \sect{env-setup}.

We study the following versions:

\begin{myitemize}
\item \naive{} incurs a round trip per register access and synchronizes entire GPU memory before/after a GPU job.

\item \cdm{}
includes selective memory synchronization~(\S\ref{sec:mem-sync}).

\item \cdmd{}, in addition to \cdm{}, includes register access deferral~(\S\ref{sec:reg-defer}); it generates per-commit round trips.

\item \cdfull{} additionally includes speculation~(\S\ref{sec:speculation}).
It represents \sys{} with all our techniques, 

\end{myitemize}

\paragraph{Recording delays}
Figure~\ref{fig:recording-delay} shows the end-to-end recording delays.
\naive{} incurs long recording delays even on WiFi
ranging from 52 seconds (MNIST, a small NN) to 423 seconds (VGG16, a large NN).
Such delays become much higher on the cellular network, range from 116 seconds to 795 seconds. 
As discussed in \sect{overview:delay}, such high delays not only slow down ML workload launch but also hurts interactivity because the TEE must lock the GPU during recording. 
Compared to \naive{}, \cdfull{} reduces the delays by up to 95\% to 18 seconds (WiFi) and 30 seconds (cellular) on average. 
We deem these delays as acceptable, as they are comparable to mobile app installation delays reported to be 10 -- 50 seconds~\cite{appTime}.

\paragraph{Replay delays}
\sys{}'s replay incurs minor overhead in workload execution as shown in Table~\ref{tab:replay}. 
Compares native executions, \sys{}'s replay delays range from 68\% lower to 3\% higher (25\% lower on average). 
\sys{} performance advantage comes from its removal of the complex GPU stack. 
We notice that these results are consistent with the prior work~\cite{tinyStack}.

\subsection{Validation of key designs}
\label{sec:eval-choice}

\paragraph{Efficacy of deferral}
As shown in Figure~\ref{fig:recording-delay} (\cdm{} vs. \cdmd{}), register access deferral reduces the overall delays by 65\% (WiFi) and 69\% (cellular). 
Table~\ref{tab:model_desc} further shows that the deferral reduces the number of round trips by 73\% on average. 
With deferral, each commit encapsulates 3.8 register accesses on average.

\begin{table}[t!]
	\centering
	\includegraphics[width=0.48\textwidth{}]{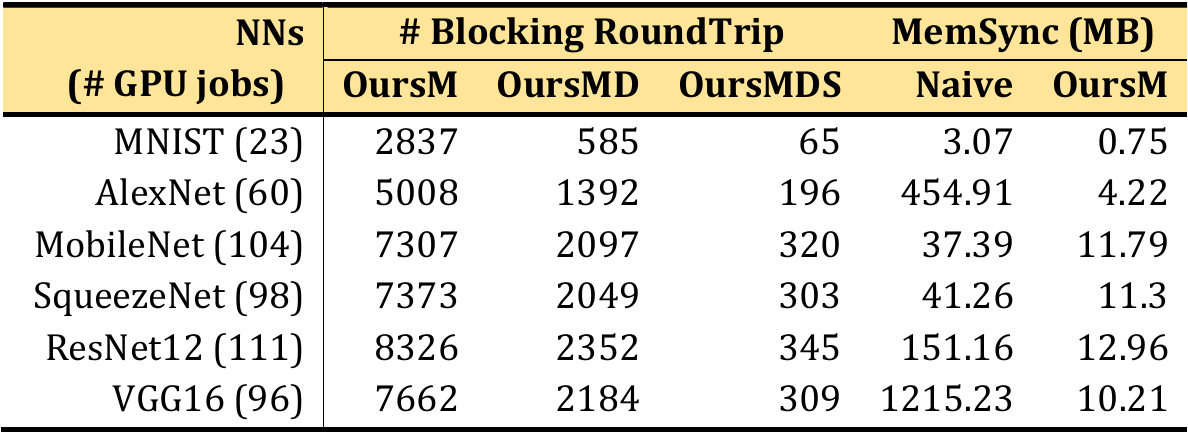} 
	\caption{Statistics of record runs, showing \sys{} significantly reduces network round trips that block the recording and the memory synchronization traffic}
	\label{tab:model_desc}
\end{table}
\paragraph{Efficacy of speculation}
We run all six benchmarks with retaining register access history in between, allowing \sys{} to reuse history across benchmarks. 
Figure~\ref{fig:recording-delay} (\cdfull{} vs. \cdmd{}) shows that speculation reduces the recording delays by 60\% to 74\%. 
Table~\ref{tab:model_desc} further shows \cdfull{} achieves 86~\% reduced number of round trips on average.
Such benefit mainly come from coalescing round trips of asynchronous commits.

We further investigate the speculation success rates and find 95\% of commits (99\% register accesses) satisfy the speculation criteria~(\S\ref{sec:speculation}). 
These commits are generated by GPU driver routines that roughly fall into four categories. 
(1) \textit{Init}: probe hardware configuration when the driver is loaded.
(2) \textit{Interrupt}: read and clear interrupt status.
(3) \textit{Power state}: periodic manipulation of GPU power states. 
(4) \textit{Polling}: busy wait for GPU to finish TLB or cache operations. 
Figure~\ref{fig:speculation} shows a breakdown of commits by category. 
All register values in these commits are highly predictable.

%
%

The commits that fail the criteria are due to reads of nondeterministic register values.
For example, on each job submission, the Mali driver reads and writes a register \texttt{LATEST\_FLUSH\_ID} which reflects the GPU cache state and can be nondeterministic.

\begin{table}[t!]
	\centering
	\includegraphics[width=0.48\textwidth{}]{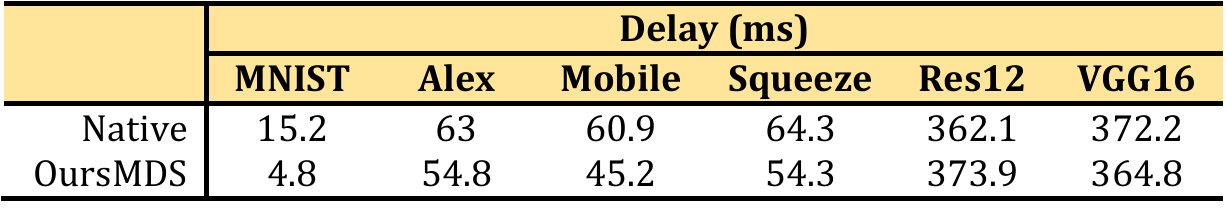} 
	\caption{Replay delays of \sys{} (\cdfull{}) are similar to Native, which executes benchmarks on the GPU stack in the normal world of the same device}
	\label{tab:replay}
	\vspace{-12pt}		
\end{table}
\paragraph{Misprediction cost}
For the above reasons, we have not observed misprediction in our 1,000 runs of each workload.
To validate that \sys{} can handle misprediction, 
we artificially inject into record runs wrong register values. 
In all the cases of injection, \sys{} always detects mismatches between the speculative and the injected register value, initiating rollback of the software and the hardware states properly.
In the worst case (misprediction at the end of a record run), we measure the delays of rollback is 1 and 3 seconds for MNIST and VGG16, respectively. 
The delays are primarily dominated by driver reload and GPU job recompilation, which overshadow the replay delays on the client GPU hardware. 

\paragraph{Selective memory synchronization}
Figure~\ref{fig:recording-delay} (\cdm{} vs. \naive{}) shows that the technique reduces the recording delays by 1 -- 57\% on average.
The reduction is more pronounced on large NNs such as AlexNet and VGG16 (34 -- 57\%). 
Table~\ref{tab:model_desc} shows the network traffic for memory synchronization is reduced by 72 -- 99\%. 




\paragraph{Polling offloading} (\S{\ref{sec:off-polling})
The numbers of polling instances range from 117 (MNIST) to 492 (VGG16),
that generate from 130 to 550 round trips.
Offloading polling reduces the total round trips by 13 -- 58, making the cost of polling instance one RTT;
This is because without offloading, a polling loop often takes a few RTTs (the RTT is long as compared to GPU operations being polled such as cache flush); 
with offloading and speculation, the RTTs often become hidden.





\begin{figure}[t]
	\centering
	\includegraphics[width=0.45\textwidth{}]{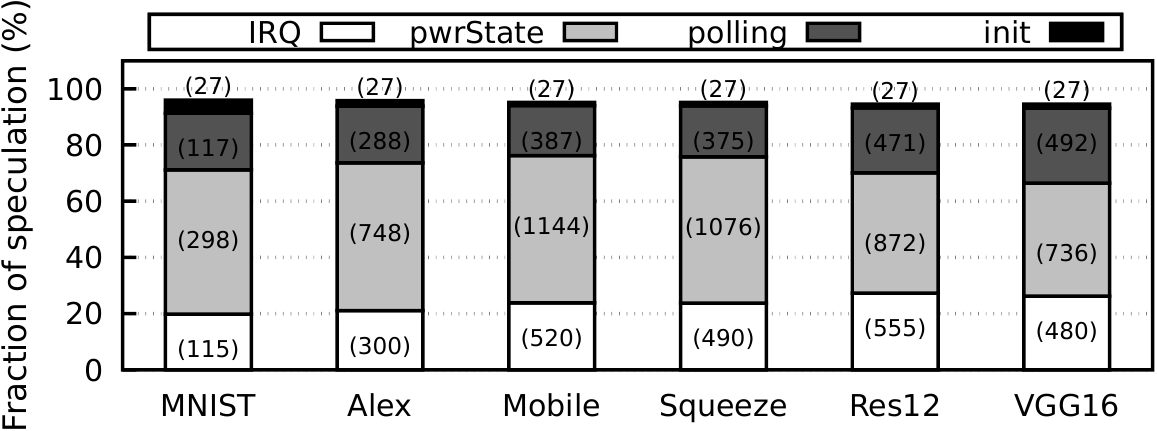} 
	\vspace{2pt}		
	\caption{Breakdown of speculation; the actual number of commits for each segment are shown in parentheses.}
	\label{fig:speculation}
	\vspace{-10pt}		
\end{figure}
\subsection{Energy consumption}
\label{sec:eval:energy}
We measure the whole client energy using a digital multimeter which instruments the power barrel of the client device (Hikey960). 
The client device has no display. 
It uses the on-board WL1835 WiFi module for communication; 
it does not run any other foreground applications. 
Each workload runs 500 iterations and we report the average energy. 
Figure~\ref{fig:energy} shows the results. 

\textit{Record.}
The energy consumed by recording is moderate, ranging from 1.8 -- 8.2 J, 
which is comparable to one by installing a mobile app, e.g. 16 J for Snapchat (80MB) on the same device.
Note that it is one-time consumption per workload.
Compared to \naive{}, \sys{} reduces the system energy consumption by 84 -- 99\%. 


\textit{Replay.}
As a reference, we measure replay energy per benchmark.
It ranges from 0.01 -- 1.3 J, consistent with the replay delays in Table~\ref{tab:replay}.
The replaying energy is comparable with the native execution on the original GPU stack of the client device (not shown in the figure).


\section{Related Work}
\label{sec:related}

\paragraph{Remote I/O}
is adopted for cross-device I/O sharing~\cite{rio,mobilePlus} and task offloading~\cite{kahawai,telekine}.
Unlike \sys{}, however, their remoting boundary is at higher-level -- device file~\cite{rio}, Android binder IPC~\cite{mobilePlus}, and runtime API~\cite{telekine}.
Such clean-cut boundaries ease a course-grained I/O remoting, e.g. function-level RPC calls.
To apply to TEE, however, the client TEE must keep a part of (e.g. device driver) or the entire I/O stack while bloating TCB.

Similar to \sys{}, prior works have explored the lowest software level;
For efficient dynamic analysis, 
they forward I/O from VM to mobile system~\cite{charm} or low-level memory access from emulator to real device~\cite{avatar,surrogates}.
However, their cross-device interfaces are wired, faster than what \sys{} addresses, (i.e. wireless connection).
\sys{} has a different goal: hosting a dryrun service for GPU recording, mitigating communication cost.

\paragraph{Device isolation with TEE}
Recent works propose TEE-based solutions for GPU isolation 
by hiding GPU stack in the TEE~\cite{HIX,secDeep} or security-critical GPU interfaces in the GPU hardware~\cite{graviton}.
They, however, require hardware modification and/or bloat TCB inside TEE.
Favorable to the insight given by \gpurip{}~\cite{tinyStack},
\sys{} offers a remote recording service for clients to reproduce GPU compute without the stack in TEE.


Leveraging TrustZone components, prior works build a trusted path locally, e.g. for secure device control~\cite{seCloak} or remotely~\cite{trustUI}, e.g. to securely display confidential text~\cite{schrodinText} and image~\cite{rushMore}.
Their techniques are well-suited to \sys{} for
i) discarding adversarial access to the GPU while recording and replaying;
ii) building secure channel between cloud VM and client TEE;


\begin{figure}[t]
	\centering
	\includegraphics[width=0.45\textwidth{}]{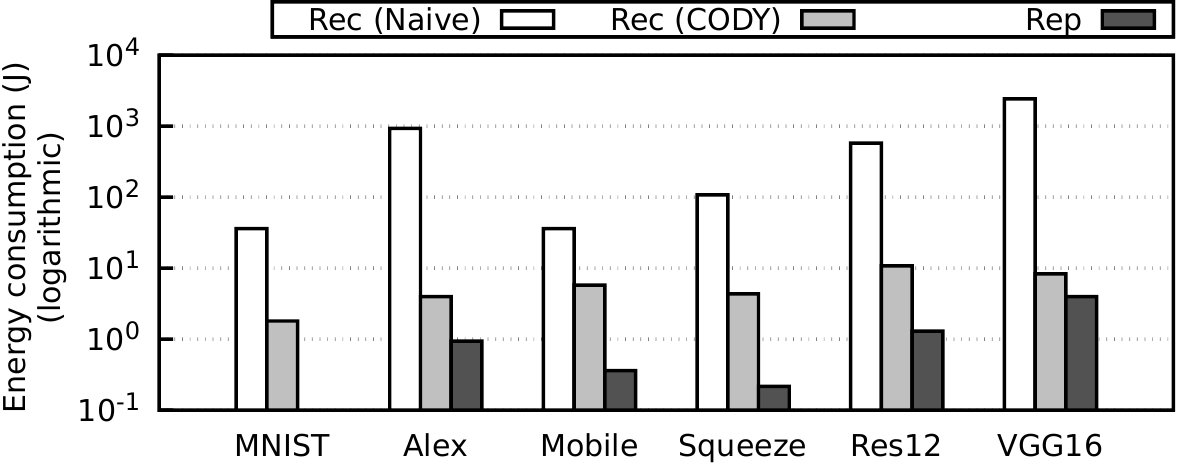} 
	\vspace{2pt}		
	\caption{Energy consumption for record and replay}
	\label{fig:energy}
	\vspace{-10pt}		
\end{figure}
\noindent
\textbf{Speculative execution}
is widely explored by prior works;
based on caching and prefetching, they facilitate asynchronous file I/O~\cite{specHint,specNFS,sprint} or speed up VM replication~\cite{remus} and distributed systems~\cite{matei08osdi}.
Unlike such works, \sys{} does not prefetch I/O access ahead of time;
instead, \sys{} hide I/O latency by speculatively continue driver's workflow deferring read values while replacing them as symbolic expression;
it then commit when facing value/control dependencies.

\paragraph{Mobile cloud offloading}
There has been previous works on cloud offloading~\cite{clonecloud,maui,comet}, which
partitions mobile application into two parts: one for local device and the other for the cloud.
Facilitating application-layer virtual machine or runtime, each part of the application cooperatively runs from both sides continuously.
Unlike them, \sys{} does not require runtime or vm support from the client;
the offloading is also temporal for dryrun of GPU compute to capture the interactions.

\noindent
\textbf{GPU record and replay}
has been explored to dig out GPU command stream semantics~\cite{rosenzweig-m1,panfrost,grate}, enhance performance~\cite{nimble}, migrate runtime calls~\cite{patrace}, and reproduce computation~\cite{tinyStack}.
While they care \textit{what} to record, \sys{}'s focus is \textit{how} to record;
\sys{} addresses costly interaction overhead for remote GPU recording.

\paragraph{Secure client ML}
Much works has been proposed to protect model and user privacy~\cite{ppfl,darkneTZ} and/or to secure ML confidentiality~\cite{ternary,occlumency}.
However, they all lack GPU-acceleration which is crucial for resource-hungry client devices.
While recent work~\cite{slalom} suggests a verifiable GPU compute with TEE, the complexity of homomorphic encryption significantly burdens client devices.

\section{Conclusions}
\sys{} provides a cloud service for GPU recording in a secure way;
it performs GPU dryrun interacting with the client GPUs over long wireless communication.
Retrofitting known I/O optimization techniques,
\sys{} significantly reduces the time and energy consumed by client to get a GPU recording.

\bibliographystyle{abbrv}
\bibliography{bib/abr-short,bib/xzl,bib/hongyu,bib/misc,bib/book,bib/security,bib/iot,bib/datacentric,bib/hp,bib/numa,bib/ml-edge,bib/secureGPU,bib/TrustZone,bib/tzgpu,bib/gpusb,bib/sgx}



\end{document}